\documentclass{article}
\usepackage{amsmath}
\pdfoutput=1 

\usepackage{array}
\usepackage{graphicx}
\usepackage[utf8]{inputenc}
\usepackage[left=1in, right=1in, top=1in, bottom=1in]{geometry}

\usepackage[english]{babel}
\usepackage{xcolor}
\usepackage{microtype}
\usepackage{tikz}

\usepackage[title]{appendix}

\providecommand{\keywords}[1]
{
  \small	
  \textbf{\textit{Keywords---}} #1
}

\title{Modelling Opaque Bilateral Market Dynamics in Financial Trading: Insights from a Multi-Agent Simulation Study}

\author{Alicia Vidler  \footnote{ A.Vidler@unsw.edu.au, School of Computer Science and Engineering, University of New South Wales, Australia}, Toby Walsh \footnote{T.Walsh@unsw.edu.au, School of Computer Science and Engineering, University of New South Wales, Australia}}

\begin{document}
\maketitle

\begin{abstract}
 Exploring complex adaptive financial trading environments through multi-agent based simulation methods presents an innovative approach within the realm of quantitative finance. Despite the dominance of multi-agent reinforcement learning approaches in financial markets with observable data, there exists a set of systematically significant financial markets that pose challenges due to their partial or obscured data availability. We, therefore, devise a multi-agent simulation approach employing small-scale meta-heuristic methods. This approach aims to represent the opaque bilateral market for Australian government bond trading, capturing the bilateral nature of bank-to-bank trading, also referred to as "over-the-counter" (OTC) trading, and commonly occurring between "market makers".
The uniqueness of the bilateral market, characterized by negotiated transactions and a limited number of agents, yields valuable insights for agent-based modelling and quantitative finance. The inherent rigidity of this market structure, which is at odds with the global proliferation of multilateral platforms and the decentralization of finance, underscores the unique insights offered by our agent-based model. We explore the implications of  market rigidity on market structure and consider the element of stability, in market design. This extends the ongoing discourse on complex financial trading environments, providing an enhanced understanding of their dynamics and implications.

\end{abstract}


\section{Introduction}

We venture into the realm of applying agent-based modelling methods to the Australian government bond market. Our approach employs single level multi-agent programming to simulate the behaviour of trading between banks, also known as "market makers". This facet represents a centralization within the existing market structure. Unlike equity markets, the Australian government bond market does not operate through a single regulated trading exchange. Instead, clients aiming to transact government bonds must interact with market makers across numerous disjointed trading platforms, electronic messaging services, and even directly with human traders via non-digital means such as via telephone.

Contrary to stock markets, where prices are primarily driven by supply and demand, the prices in government bond markets are significantly impacted by the respective countries' central banks through their management of interest rate policies and, more recently, inflation rate expectations. Although markets can price in the potential for changes in interest rates, bond prices largely reflect the current rate-setting policy.

Market participants typically concentrate on facilitating bond sales and purchases, rather than the trading price of the bond. This emphasis is partially explained by the contractual interest payments associated with bonds. By owning a bond, the holder receives returns (interest payments), making the bond market unusual in that prices are less critical than access to "flow" or "liquidity". Our work primarily focuses on modelling this "flow" or the movement of bonds through the market.

Additionally, bond markets have a number of features which are not relevant for our modeling purpose but are worth highlighting.  The price of a bond is inversely related to the yield that the current market expects.  As part of a governments bond issuance program, the mechanism through which they borrow funds, a number of bonds will be issues with different maturity rates.  Our work is unaffected by these features as we look to simulate a overall market design concept of liquidity transfer and government bonds are largely substitute for different maturities.  In Australia this is typically the 5 year bond.

Considering the few market makers (four) compared to the vast number of bond buyers and sellers (potentially over 2000), alongside the lack of direct observability of markets and agents, calibrating parameters of an agent-based model becomes complex. Despite no consensus method for this \cite{avegliano2019using}, we propose an inductive approach using market structure data and \cite{Pinter2023} to specifying agent calibrations. The utility functions and policies utilised within our simulation approach stem from finance literature and regulatory requirements. We apply agent-based model simulation techniques to small populations of heterogeneous agents, analysing market stability and agent trading features. Our modelling reveals crucial insights into the role these market maker agents play in bond market participant interactions, providing implications for the future regulation of this marketplace.

We simulate heterogeneous market makers trading a single type of product: a 5-year Australian government bond \footnote{The AOFM reports the current duration of bonds on issue to be 5.9 years, closely matching the 5-year profile}. Each market maker is modelled as an individual agent with its own adaptive utility function. Agents acquire bonds from their unique, non-overlapping client base, as characterised by the Australian Office of Financial Management (AOFM). This client base includes various sectors, such as Interbank, Bank balance sheet, other domestic, Japan, Asia (excluding Japan), UK, Europe, Americas, and other international. Since the initiation of data release in 2016, Interbank has consistently represented the single largest share of secondary market turnover, averaging 26\% of all reported volumes as trades between themselves.

\section{Background: Bond Markets and Agents}
Global bond markets encompass diverse participant types or agents. In Australia, these include major retail banks acting as trading facilitators (market makers), superannuation funds as typical debt purchasers, and hedge funds or other trading houses functioning as clients. Beyond participants, several markets or platforms host bond trading, analogous to a stock exchange but for bonds. Notably, Australia lacks a dominant trading venue, resulting in a highly fragmented bond market dispersed across numerous private software platforms. This landscape contrasts starkly with equity markets that have a unified trading mechanism under a single regulatory authority. The bond markets, conversely, are opaque and decentralised.

Following the COVID-19 pandemic, the Australian Federal Government augmented its borrowing to fuel a variety of social initiatives and direct economic support measures. As of April 30, 2023, the Commonwealth of Australia's gross debt on issuance stands at \$894 billion, with \$300 billion held by the Reserve Bank due to quantitative easing \footnote{https://www.afr.com/markets/debt-markets/reserve-bank-holds-off-on-selling-bonds-20230524-p5db0w}. The Australian Office of Financial Management (AOFM) \footnote{https://www.aofm.gov.au/data-hub} collates and releases only high-level macro trading data, reporting that domestic market makers conducted 28\% of all bond trading transactions in 2022. Additionally, the Australian Securities and Investments Commission oversees the licensing of exchange mechanisms and regulates the behaviour across these platforms \footnote{Regulation pertinent can be found here RG 172 https://download.asic.gov.au/media/okhpmov0/rg172-published-2-august-2022.pdf}.

In our quest for viable modelling approaches and AI-based techniques, it became evident that the complexity of the bond market necessitates a method capable of modelling nonlinear dynamics with limited data. Notably, intricate interactions characterize this dynamic system, and these cannot be simplified readily. We delve into a historical overview of the explored techniques and approaches to address this challenge. Bond market design research remains a nascent field, with much of the focus placed on bond pricing models and inferring default probabilities rather than the market's design and structure. A significant portion of bonds continues to be traded "over-the-counter". As global markets transition to electronic trading, the resultant transparency begins to resonate with areas of interest in \cite{Asquith_2013}. Regulatory demands for increased transparency \cite{Asquith_2013}, along with growing concerns about bond market structures \cite{Owens_2018}, have invoked considerations of regulatory changes' potential impacts on future market designs \cite{Blommestein_2017}. The ongoing market evolution, however, seems hindered by data scarcity, inadvertently impeding academic literature in this realm (see \cite{Aman_2019}).

\subsection{Market micro structure and trading venues}

Bond markets, characterized by their opacity and bilateral nature, predominantly transact through software platforms controlled by a handful of firms like TradeWeb, Bloomberg, LiquidityCube, ICAP, MarketAxess, LiquidNet, and Yeildbroker. These platforms act as facilitators, essential for agent trading without a clearing bank or similar entity. In May 2023, significant industry consolidation was announced. TradeWeb acquired Yeildbroker \footnote{https://www.afr.com/chanticleer/bond-market-players-call-time-on-6-trillion-aussie-exchange-20230525-p5db5n}, which dominates the Australian bond market, while ICAP purchased LiquidNet. Additionally, Bloomberg, TradeWeb, and MarketAxess launched a joint venture centralizing bond market transaction data collection \footnote{https://www.tradeweb.com/bloomberg-marketaxess-and-tradeweb-sign-joint-venture-agreement}. This reshuffling, driven partly by regulatory demands for transaction visibility, prompts our work to offer a fresh modeling approach to the evolving market microstructure.

\subsection{Previous work on Market mechanisms}

Mechanism design, a subset of Game theory, is often employed to analyze decentralized systems and their attributes. Such research is often "used as a tool to analyze decentralised systems and their properties"  \cite{Mansour_2014}. It explores the creation of mechanisms that implement a social choice function, considering rational participant behavior. With advancements, it's now possible to approximate sub-optimal algorithmic design principles. Further work by \cite{Dobzinski_2013}, \cite{Lee_2019}, \cite{Procaccia_2013}, \cite{Donna_2020}, \cite{Polanski_2017} and \cite{Cai_2020} also covers this. However, these markets fundamentally differ from the bond market we look to model. Bond markets aren't composed solely of price-taking agents, and there's no evidence suggesting a shift in demand for collaborative purposes.

% The auction process in traditional asset markets, like equities, is characterized by Walrasian mechanisms and efficiency \cite{DBLP:journals/corr/BabaioffLNL13}.Other similar work in this direction include \cite{Dobzinski_2013}, \cite{Lee_2019}, \cite{Procaccia_2013}, \cite{Donna_2020}, \cite{Polanski_2017} and \cite{Cai_2020}. However, these markets fundamentally differ from the bond market we look to model. Bond markets aren't composed solely of price-taking agents, and there's no evidence suggesting a shift in demand for collaborative purposes.

% There is interesting research on the application of mechanism design to unique marketplaces \cite{DBLP:journals/corr/Gonczarowski13} , investigating how market outcomes can be manipulated through design and collaboration. Within financial markets this implies that mechanism design principles can affect outcomes that begin to approach market manipulation.  

\subsection{Agent-Based Models}
Agent-based modeling (ABM), employing computational methods to depict various agents' interactions, is aptly suited to elucidating bond market dynamics. This approach's strength lies in simulating scenarios involving interdependent and adaptive agents, facilitating the reproduction of complex dynamic systems \cite{Bai_2020, Gilbert_2007}. Further exploration into ABM is detailed in \cite{Fagiolo_2017, Lussange_2020, Vermeir_2015}. We aim to extend this field by applying ABM to a partly decentralized marketplace characterized by abundant goods sources and relatively few agents, integrating the theory of Adaptive Market Hypothesis (AMH) introduced in \cite{Lo_2004}. AMH investigates the interaction between market participant behaviour and design features \cite{Lo_2004}, aligning well with ABM analysis.

\subsubsection{MESA and SugarScape: ABM for Complex Dynamic Systems}
SugarScape signifies the application of ABM to the growth of artificial complex societies, pioneered by \cite{Axtell}. This simulation mirrors a society where agents and their environment interact following a set of rules. The model starts with few rules applied to specific agents, then evolves to examine the resulting marketplace. It features various extensions like agents' coalition formation, meta structures, and trading \footnote{Mesa is available on GitHub at https://github.com/projectmesa/mesa}. Notable iterations of the \cite{Axtell} framework and its MESA extension are examined in \cite{Pike_2019, Masad_2015, Pike_2021, Kazil_2020}.

\subsection{Agent-Based Models Employing Reinforcement Learning}
Recent advancements in reinforcementf learning have incorporated agent models for training \cite{vadori2022multiagent}. This approach largely reduces market participants into two categories concerning liquidity provision—either providers or consumers. Echoing the work by \cite{BROCK19981235}, the authors propose a game-theoretic approach employing two agent categories to model all participants' utility function. The model shows potential in deducing latent behaviours and utility functions of the two agent types by repeatedly playing "games" using market data.

Our work, representing bond market trading banks, demonstrates a preference for agents to avoid inter-trading to clear inventory, typically driven by an asset imbalance influenced by client demand. This utility function significantly differs from the binary categorization of agents into either liquidity providers or consumers (see appendix).

As previously discussed, no consensus exists on the best method to calibrate an ABM \cite{avegliano2019using}. The complexity of environments we aim to model often precludes the observation of agents' utility functions, policies, or response functions \cite{boggess2022toward}. Machine learning has been applied to estimate the weights, calibrations, policies, and other factors within an ABM across diverse economic or social systems \cite{Ardon2021}. In the finance domain, this has proven successful in the foreign exchange trading market, where models rely on finding calibration sets through reinforcement learning \cite{ganesh2019reinforcement, vadori2022multiagent}. However, these market domains substantially differ from the bilateral bond market we aim to model. This existing body of work often benefits from features absent in our target market, including the reduction of agents into only two classes, homogeneous in utility function and form, and mutually exclusive (Liquidity Providers or Liquidity Takers, as defined in \cite{vadori2022multiagent}). Additionally, previous work depends on either a highly observable market environment or a unique environment description due to regulatory requirements or other functions \cite{Banks2021, barzykin2021algorithmic, cont2021stochastic}. Employing machine learning techniques (such as reinforcement learning), researchers create a "surrogate" model. While surrogates can improve computational efficiency by reducing parameter dimensionality \cite{DeepUQ2018}, they do not aid in initial parameter calibration \cite{avegliano2019using}. 

We contribute to this body of work by modelling a less advanced yet more challenging market dynamic, exploring an inductive approach to calibration.

\subsection{Bond market research}
Bonds literature predominantly centers around pricing products \cite{Longstaff_1995},\cite{Duffie_1999}, \cite{Black_1976}, \cite{Merton_1974}, \cite{Liang_2017} and advanced alternative pricing methods \cite{AlbaneseVidler, Albanesedynamic, Albanese06}. Limited studies have explored bond market design, often veering towards related markets like corporate bonds \cite{Braun-Munzinger2018} or bond futures \cite{Sun_2018}. Notably, the work of the U.S. Securities Exchange Commission \cite{SEC} and Lin et al. \cite{Lin_2020} applying agent-based modeling to equity markets and corporate bond markets respectively, aligns with our research. We aim to extend these methodologies to the Australian bond market's unique dynamics. The universally applicable agent-based modelling approach, evidenced by \cite{Leal_2014, Fagiolo_2017, Lussange_2020, Vermeir_2015}, is utilized in our work for a specific application. This allows us to explore market topology \cite{Todd}, \cite{Bonabeau_2002} and dynamics \cite{Braun-Munzinger2018}, \cite{Lin_2020}, \cite{Cockburn_2010}, \cite{Hui_2010}, \cite{Qian_2016} within the context of the Australian government bond trades regulation.

\subsection{Validation of ABM's}

Agent-based models (ABMs) lack a universally agreed-upon calibration approach, prompting us to seek similarities with ABM research in other technical sectors, including applied system dynamic theory and social constructed mental models. The "element distance ratio" is utilized to measure distance between two model structures, assessing each model's credibility. This concept has been applied to ABMs in previous research, comparing an inductively calibrated ABM to a deductively calibrated one.

In our research, which involves inductively calibrated ABMs, we resort to sensitivity analysis for conceptual model validation. The validation method of comparing the same input data in two differently calibrated models and assessing the distance ratio is employed, with a smaller ratio indicating closer model alignment. Extensive discussions around the challenges of such methods exits in \cite{quera2023some}

We also consider the five "value-based" validation criteria derived from systems design work, which entail expert consultation on five validity points. However, in a market where participants are bound by non-disclosure agreements, such consultations involving subjective judgments may not be fully applicable. Recognising the importance in user trust, explainable systems work exists though presents more challenges than solutions \cite{suryanarayana2022explainability}

Surrogate models—low-dimension models with fewer parameters—are examined as a recent advancement in the field. Heterogeneous agent-based models (HAMs), as explained in the research, are able to effectively replicate financial data series' stylized facts. These models also demonstrate their capability in explaining market anomalies like bubbles, crashes, and sources of market "chaos". However, surrogate models tend to be built on markets that adhere more to normality or distributional assumptions and are often trained via multiple simulated games based on simulated market pricing.

It's important to note the reliance of these methods on large training data sets, and the lack of a definitive way to ascertain if a data set is sufficiently large or comprehensive for robustness. While much of the research focuses on asset pricing with HAMs, our work applies these techniques to the market's underlying structure. Unfortunately, these HAM-based techniques aren't applicable in our case, leading us to depend on sensitive analysis and matching historical data, all while adhering to regulatory requirements.

\section{Method}

The proposed inductive approach utilizes market structure data from the Australian Government's Office of Financial Management and survey results for agent calibration.  See appendix for specific model calibration is details. The simulation incorporates utility functions and policies sourced from finance literature and regulatory mandates.  We utilise the power and complexity of the Sugarscape model as its naturally holds many similarities to the human interaction methods of a bilateral market. For more detail on the design of Sugarscape see \cite{Axtell} and \cite{Axtell2022} .

\subsection{Agent Interaction and Trading Mechanism}

The model investigates the dynamic interaction of agents, trading bilaterally. The bilateral trading mechanism is medium-agnostic, implying trades can occur through various channels. This model extends the framework of Axtell and Epstein, applying the features of the market maker dynamic to the Sugarscape formalism. The synergies between these frameworks manifest as follows:

\begin{enumerate}
    \item Bond collection by agents from clients is labelled "collecting," similar to the term used in \cite{Axtell}.
    \item Market makers need bonds for funding purposes, analogous to \cite{Axtell}'s concept of "metabolism".
    \item Each agent has a different and overlapping client base, akin to "vision".
    \item By employing "metabolism", agents can quantify their bond accumulation desire, called the "marginal rate of substitution".
    \item The "marginal rate of substitution" is a metric derived from each agent's ability to acquire bonds and cash, balanced against their internalized asset usage.
    \item Trading periods are standardized as \(t \in I[0,t]\).
\end{enumerate}

%%%%%%%%%%%% flow diagram

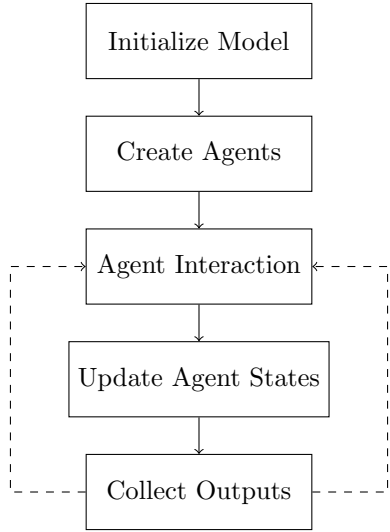
\begin{figure}[htbp]
\centering
\begin{tikzpicture}[node distance= 1.5cm, every node/.style={draw, rectangle, minimum width=3cm, minimum height=1cm, align=center}]

% Nodes
\node (init) {Initialize Model};
\node (agents) [below of=init] {Create Agents};
\node (interaction) [below of=agents] {Agent Interaction};
\node (update) [below of=interaction] {Update Agent States};
\node (output) [below of=update] {Collect Outputs};

% Arrows
\draw [->] (init) -- (agents);
\draw [->] (agents) -- (interaction);
\draw [->] (interaction) -- (update);
\draw [->] (update) -- (output);

% Optional arrows for iterative process
\draw [->, dashed] (output.west) -- ++(-1,0) |- (interaction.west);
\draw [->, dashed] (output.east) -- ++(1,0) |- (interaction.east);

\end{tikzpicture}
\caption{Agent-Based Model Workflow}
\label{fig:abm_workflow}
\end{figure}

%%%%%%%%%% end flow diagram

\subsection{Market Characteristics}

The proposed agent-based model represents the current trading environment with the following features: 

\begin{enumerate}
  \item \textbf{Agents}: "Market makers," majorly from the four Australian retail banks, collect assets from clients and trade with other market makers as a "last resort".
  \item \textbf{Environment}: The origin of Agents, also determining the client flow and inventory.
  \item \textbf{Market Rules}: The strategies agents use for interaction, though not always government regulated.
\end{enumerate}

\subsection{Agent and Landscape Characteristics}

Agents, who are buyers, sellers, or traders, possess inherent needs. They locate resources, transact with others in their "vision," and can be removed after fulfilling their resource objectives. The marketplace, populated with resources held by clients, offers agents the following actions:

\begin{enumerate}
    \item Collect resources
    \item Consume resources
    \item Cease business operations
    \item Trade with other agents if a utility valuation difference exists
\end{enumerate}

A crucial marketplace feature is "vision," which details an agent's access to client resources. In the financial industry, this is known as "franchise value" or client relationship depth.

\subsubsection{Policies}

A simple policy set is calibrated for all agents, where the model design triggers a trade for a welfare-improving function. The welfare function evolves over time as agents accumulate and use resources. This reflects actual trading environments for government bonds, creating a wealth effect for older books. Different participants can trade with other agents at various prices, with no clearing price established. The model assumes a "welfare-improving" policy for bilateral trade between agents, mimicking actual market operations.

\section{Code Implementation}

Leveraging the crucial role of heterogeneous agent beliefs in financial markets, as highlighted by \cite{BROCK19981235}, we employ various open-source code bases for implementing a customized version of "SugarScape" that simulates the regulated market structure of bond markets.  We assume a discrete time approach, with asynchronius agent actions. Notably, \cite{SpecofSugar} provides a comprehensive agent-based model framework established by Sugarscape and commercially developed software. Our implementation leverages open-source code from \cite{python-mesa-2020}, along with libraries from \cite{SugarScape} and \cite{MultiMesa}. 

We employ the following steps(see appendix for further details)

\begin{enumerate}
    \item Initiate the simulation with four agents (matching the number of Australian market making banks).
    \item Endow agents with predefined characteristics:
        \begin{itemize}
          \item Vision, representing client coverage breadth 
          \item Metabolism, indicating the agent's "carrying cost" or the cost to stay in business.
          \item Initial bond resources (akin to each trading firms intial daily position)
        \end{itemize} 
    \item Assign each agent a randomly determined starting territory or client base at inception of every simulation. 
    \item Enable trades when there are differences in agent valuations, with clear buyer-seller distinctions. 
    \item Conduct each interaction simulation asynchronously over time steps or epochs
    \item Retain the same genetic factors at the start of each simulation, irrespective of the number of runs.
    \item Repeat each basic market design simulation up to 200 times
    
\end{enumerate}

In simulating bond markets, we integrated sources like survey responses, regulatory authority documentation, and internal resources. The concept of \textbf{stability} employed here is a system that:
\begin{enumerate}
    \item Aligns with regulatory requirements and publicly reported micro-structures.
    \item Produces similar outcomes with small changes in inputs. 
    \item Approximates trade between agents to reported data for the AOFM.
    \item Possesses a stable agent count without substantial system collapse. 
\end{enumerate}

Maintaining market functionality stability is not only vital for societal functions but is also essential for regulatory bodies. Balancing client service and inter-agent trade is fundamental to ensure the network's viability. The primary metric under focus in this work is the percentage of agent interactions that involved trading with other agents.

\section{Hypothesis: Projected Outcomes with Emphasis on Market Trading Stability}

Following our comprehensive analysis, we hypothesize the emergence of four notable outcomes. These outcomes, significantly influenced by the stability of market trading, are anticipated to carry considerable relevance not only for regulatory bodies but also in shaping and upholding the principles of fair market practices in financial trading.

% 
%%%%%%%%%%%%%%%%  OLD 

%%%%%%%%%%%%%%%%%%%

\subsection{Hypothesis 1: Role of Agent Diversity Versus Number in Market Stability}
\paragraph{Influence of homogeneous agents in system stability} Our hypothesis suggests that merely increasing the number of agents, while keeping their features homogeneous, does not contribute to market stability. The examination of systems with identical agent and landscape feature sets, varying only in the number of agents, indicates a strong correlation between homogeneity and fragility. Systems with homogeneous agent features exhibit accelerated instability, collapsing by the 25th time step, as seen in our simulations that were designed to run until the 4000th time step.

%Death by timestep diff agents, same M and V fixed.png 
\begin{figure}[ht]
  \centering
   \includegraphics[width=\linewidth]{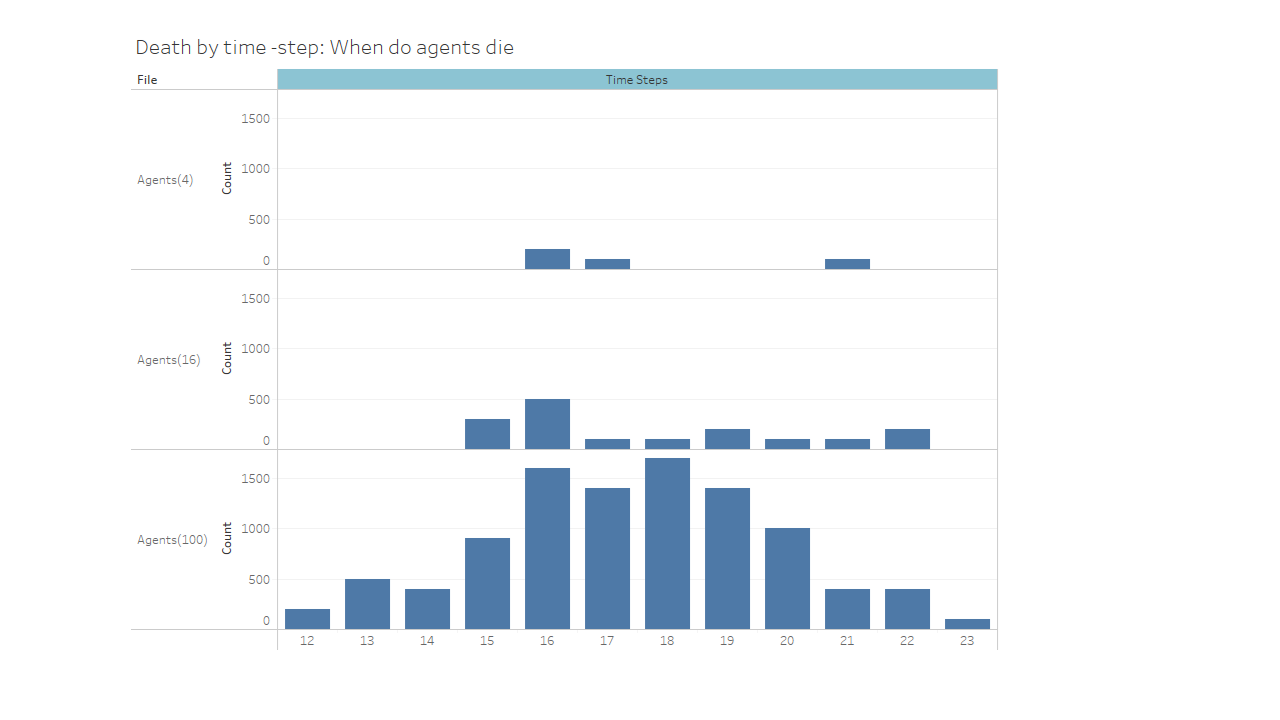}
   \caption{Impact of Homogeneity: System instability observed in populations with identical agent features}
\end{figure}

\paragraph{Implications in trading} Considering the sparse data points available from the real-world opaque marketplace, we observed the turnover of bonds between diverse clients in the Australian bond market and the number of market-making banks (represented as agents in our model). The data indicates that approximately 28\% of all turnover occurs between market-making banks ("agents"), while the remaining takes place between clients and agents in the environment. 

We further expanded our analysis by conducting additional tests and simulations. Different epochs with varying numbers of agents were simulated, attributing a range of outcomes to the distinct feature sets (such as metabolism, vision, etc). Our hypothesis underlines the essential role of agent diversity, rather than sheer numbers, in promoting the stability of market trading.

\begin{enumerate}

\item A simple test of lowly diverse agents (4) with vision and metabolism ranging between 1-5 units, trades occur less than 1\% of the time.  Broadening out vision (1-20 units) and metabolism to range between 1- 20, but still looking at 4 agents, trading occurs in 3.4\% of actions, whilst keeping metabolism lower at 1-5 units, but allowing vision to range up to 20 units produces trading of 9.2\% of interactions

% <1%  
%C:\Users\alici\Dropbox\ALICIA_2022\PhD\Bonds\MESA _ Sugarscape\SugarScape_30\brute initial _ tests\ICAIF_Low diversity\A4, v1-5, M1-5
% ~3.4%  
%C:\Users\alici\Dropbox\ALICIA_2022\PhD\Bonds\MESA _ Sugarscape\SugarScape_30\brute initial _ tests\ICAIF_Low diversity\A4, v1-20, M1-20
% 9.2%  
%C:\Users\alici\Dropbox\ALICIA_2022\PhD\Bonds\MESA _ Sugarscape\SugarScape_30\brute initial _ tests\ICAIF_Low diversity\A4, v1-20, M1-5

% wILL USE COUNT OF TRADE FLAGS 
%5.1 How often do agents trade with different vision met.png
\begin{figure}[ht]
   \centering
   \includegraphics[width=\linewidth]{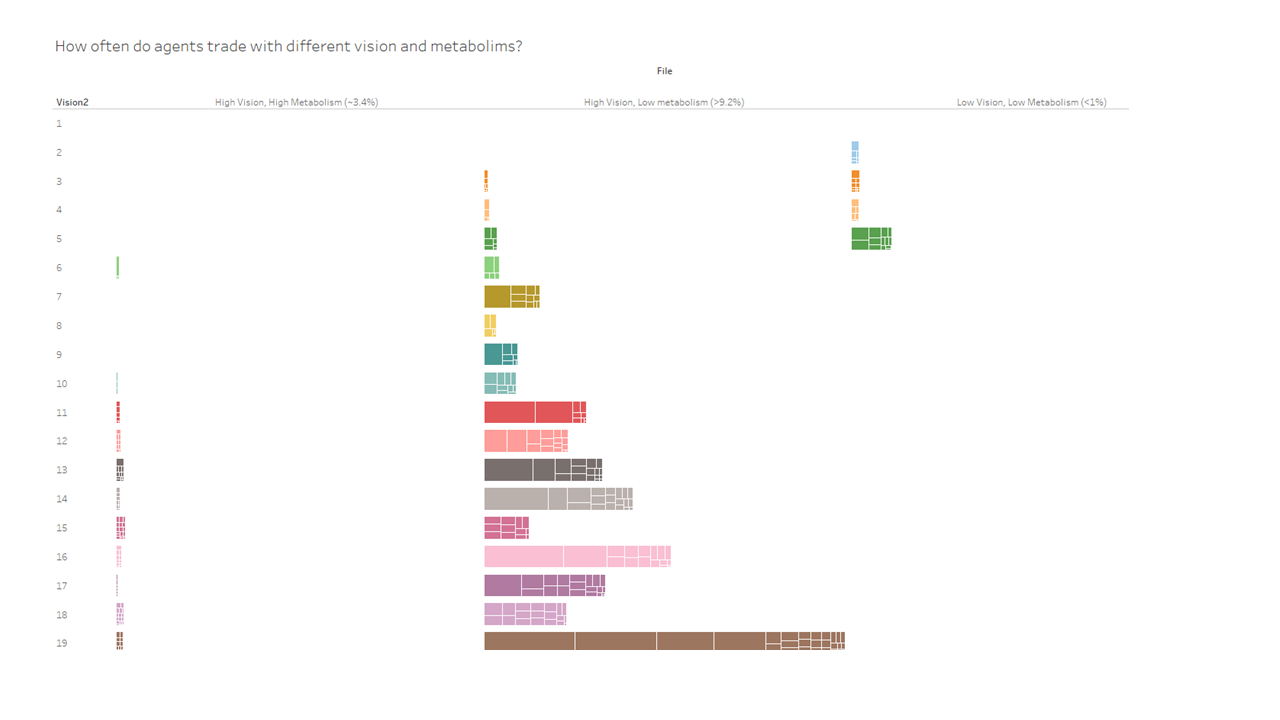}
   \caption{Trading propensity with systems of 4 agents with increasing degrees of freedom}
%   \Description{A woman and a girl in white dresses sit in an open car.}
\end{figure}

\item  Increasing agents to 16 (the current number in the UK markets) again with vision and metabolism ranging between 1-5, trades occur 6.418\% of the time 
% C:\Users\alici\Dropbox\ALICIA_2022\PhD\Bonds\MESA _ Sugarscape\SugarScape_30\brute initial _ tests\A16, v1-5,    Tinitial_Brute_r100_a16_s1000.xls

\item Increasing agents further to 100 (the approximate number of the US market) with vision and metabolism ranging between 1 -20 for each field, 38.2\% of actions resulted in trades
% C:\Users\alici\Dropbox\ALICIA_2022\PhD\Bonds\MESA _ Sugarscape\SugarScape_30\brute initial _ tests\A100, V1-20, M1-20,  Tinitial_Brute_r100_a100_v20_M20_s1000.xls
\end{enumerate}

Summarizing the findings, it can be deduced that increasing agent diversity tends to augment trading frequencies, particularly when agent populations remain relatively low. Thus, our first hypothesis supports the notion that diversity of agents plays a more crucial role in trading dynamics and market stability than simply the number of agents.

\subsection{Hypothesis 2: Easing Trading Restrictions and Reducing Agent Costs Enhances Trading Stability and Market Longevity}

An exploration of systems comprised of a small agent population, i.e., four agents, with a moderate vision range (1-10 units) but exceptionally low metabolism, yields notable results in terms of trading data. In this setup, trading activities represent approximately 9.97\% of all agent actions, marking a steady increase from the results obtained in prior analyses for 4-agent systems.

A key mechanism for imposing trade restrictions or creating barriers is the introduction of a 'metabolism' penalty function for agents. Each time step or turn in a simulation penalizes an agent by deducting a fixed amount from their accrued resource values, akin to their metabolism. In extreme instances, it is possible to evaluate the stability of a dynamic complex system by allowing each client resource area to regrow, essentially enabling resource regeneration and the potential for endless interactions due to the perpetual regrowth of resources. With low metabolism, a continuous supply of resources, and a wide vision, trading activities occur remarkably 38.7\% of the time between the four agents in this non-realistic edge case.

This outcome subtly suggests the counter-intuitive hypothesis that regulators might facilitate a more stable and liquid market by reducing trading barriers for agents. Easing restrictions, lowering costs, and ensuring a continuous supply of resources could potentially encourage more frequent interactions, thereby fostering market stability and longevity.

\subsection{Hypothesis 3: Broadening Client Base Does Not Necessarily Enhance Agent Welfare}

The intricate dynamics between agents, their feature sets, and the environment can be explored by considering an agent's 'vision' as a proxy for the breadth of their client base, while holding all other variables constant. This is pertinent to financial markets, where banks and financial institutions often aim for economies of scale, and the emergence of "fintech disrupters" introduces a wider mix of agents with diverse client bases. Furthermore, regulatory strategies that promote competition often encourage the entry of smaller, newer market players, which inherently have a narrower client base compared to established participants. These scenarios can be evaluated by investigating the impact of an agent's 'vision' on various agent designs.

From the analyses, it is evident that for systems with small agent populations, expanding the breadth of the client base or increasing 'vision' is not a universal remedy for enhancing stability, trading, or longevity. Instead, these results appear to be more influenced by the agents' 'metabolism,' or essentially, the barriers to entry and sustaining an agent in the field. 

In the context of financial markets, this is a crucial finding. Specifically, adding an agent with limited 'vision' can still lead to successful outcomes provided the agent possesses the right supportive features. However, such success scenarios aren't common. In most simulations, agents with broader 'vision' or a larger client base seem to have an 'easier' time trading with others. Interestingly, these effects can reverse if higher agent costs are introduced. 

This suggests a nonlinear relationship between trading barriers (represented by increasing metabolism or agent costs) and client base expansion. While reducing agent costs directly improves agent welfare (measured as the number of trades and time to cessation), expanding client bases usually enhances welfare, but not when costs also increase. Hence, broadening the client base does not inherently guarantee improved agent welfare.

%%% AV look at longest living agent  - USE STATS from speadsheet 

%  C:\Users\alici\Dropbox\ALICIA_2022\PhD\Bonds\MESA _ Sugarscape\SugarScape_30\brute initial _ tests\d_A4, v1-5

% wILL USE COUNT OF TRADE FLAGS 
% small vision 4 agents (V1-5) Brutal Initial.png
\begin{figure}[ht]
   \centering
   \includegraphics[width=\linewidth]{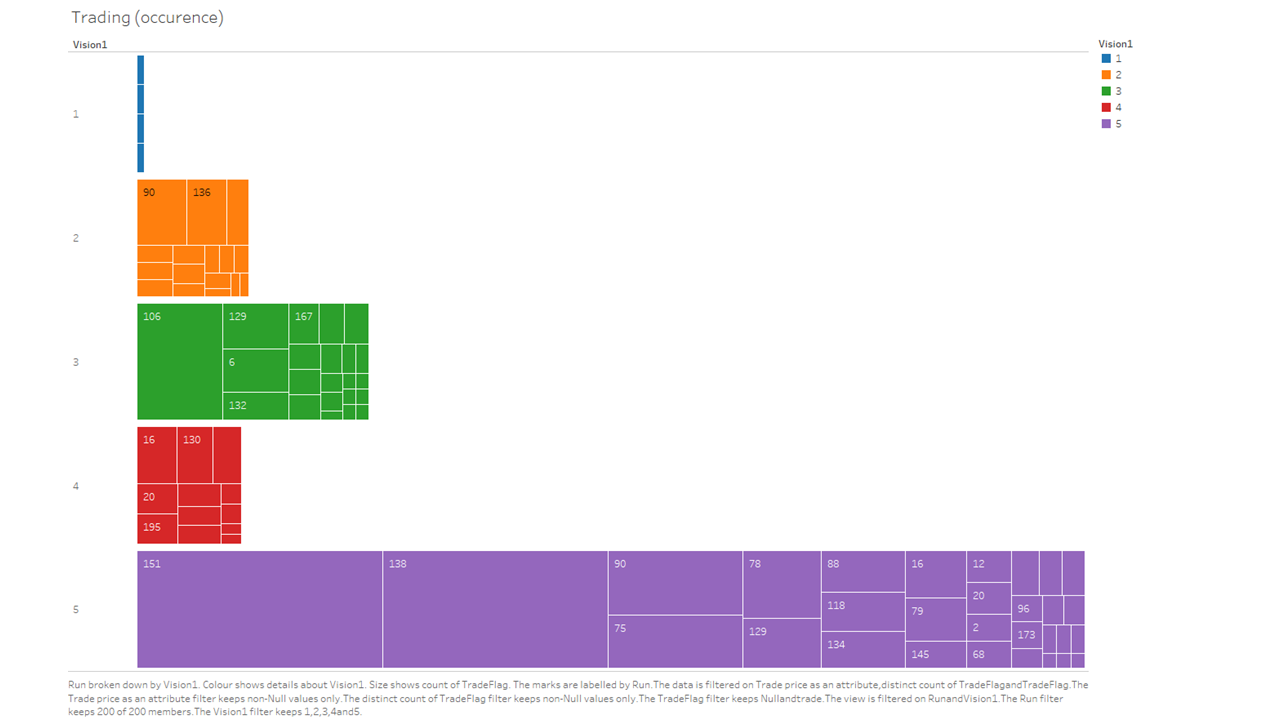}
   \caption{Agents with higher vision trade more, and with other agents of higher vision}
%   \Description{A woman and a girl in white dresses sit in an open car.}
\end{figure}

%%%%%%%%%%%%%% FILE %%%%%%%%%%%
% % C:\Users\alici\Dropbox\ALICIA_2022\PhD\Bonds\MESA _ Sugarscape\SugarScape_30\brute initial _ tests\d_A4, v1-50,    ANALYSIS_Meet calib A4_s4,000_V1-50_met1-5.xlsx

\subsection{Hypothesis 4: Is the current Market configuration the Goldilocks version? }

Our study successfully parallels an agent-based model (ABM) with the currently observed attributes of the Australian bond markets across 100 epochs, providing insights into optimal market structures. We primarily leveraged data from the Australian Office of Financial Management (AOFM), which reported interbank trading bonds turnover accounting for 28\% of traded volumes.  We test the hypothesis that the current mechanism supports the "Goldilocks" market design, i.e. that the market design is not too rigidity or fragile, and has just enough diversity for agent behaviour. 

We simulated epochs with four agents and a vast client base, represented by a vision of 50 units. This value signifies the ecosystem's maximum view from any point considering the Moore method's application over the Von Neumann method. These simulations revealed an average trading occurrence of 29.1\% of interactions, implying that agents possess overlapping client bases and complete vision of the entire environment. It's worth noting that these findings are premised on comparing AOFM's reported trading volumes with our model's trading frequency. Although it's not an exact match, it serves as a reasonable proxy in such an opaque market.

In addition, our results demonstrate that the median frequency of trading within an average simulation stands at 23.7\% of actions, highlighting considerable variance and distribution in outcomes. Nevertheless, the range of trading interactions is broad, with some four-agent simulations documenting trades as low as 2\% of the time, while others peak at 94\%. Even though averages may be skewed, the distribution reveals significant variability, further supporting the non-linear dynamics in agent interactions.

Overall, our findings provide evidence towards a 'Goldilocks' scenario—suggesting that the optimal market structure might be one that closely aligns with the current configuration of the Australian bond market.

%%%%%%%%%%%%%  CHAOS and 

\subsection{Summary}
This study presents four significant hypotheses concerning the characteristics of agent behavior, their influence on trading stability, and their role in shaping the overall structure of the financial markets, specifically focusing on the Australian bond market.

The first hypothesis illustrates the relationship between agent diversity and trading frequency, finding that greater diversity among agents leads to an increased rate of trades, suggesting that agent homogeneity, as opposed to sheer numbers, has a more considerable impact on the frequency of trades.

The second hypothesis examines the effects of reduced trading restrictions and agent costs on the market. It implies that these reductions could promote both trading stability and longevity. In a seemingly counterintuitive finding, it suggests that by easing constraints on agents, market regulators could potentially encourage a more resilient and liquid market.

The third hypothesis questions the assumption that a broader client base necessarily benefits the agent. It contends that in the complex interplay of agent characteristics and dynamics, such factors as barriers to entry and the maintenance costs associated with market presence can significantly influence outcomes, rather than the breadth of the client base alone.

The fourth and final hypothesis, often referred to as the "Goldilocks" hypothesis, aligns the agent-based model with the existing characteristics of the Australian bond market. It underscores that trading interactions exhibit considerable variation, echoing the chaotic nature of real-world markets. This observation is supported by existing literature on market behavior. On average, the trading frequency in the model aligns closely with reported trading volumes, which serves as a reasonable approximation for such an opaque market. There is a need for further research to ascertain whether these factors contribute to the market "chaos" documented in literature \cite{BROCK19981235} and \cite{Ingladaperez2020}.

This complexity and chaotic variation underscore the intricate interplay between agent costs, the breadth of client bases, and their combined effects on market stability and agent welfare.

\section{Conclusion}
%%%%%%% 2nd conclusions

This research effectively leverages agent-based modeling to depict the dynamics of the Australian bond market, focusing on market makers. Our model, informed by government data, encapsulates four agents, aligning with macro-level bond turnover data from the Australian Office of Financial Management. Through a sensitivity analysis, we studied the impacts of the number of agents, breadth of client base, and costs on market dynamics. Key insights include the benefits of agent diversity for market stability and the surprising enhancement of stability through lowered agent costs. Moreover, we found that a wide client base doesn't necessarily secure an agent's success. Even with identical initial inputs mirroring the current macro-level market structure, the system can exhibit considerable variations in stability, underscoring the chaotic nature of financial markets.

The implications for regulators are profound, offering a means to dissect the intricacies of financial markets including centralization, client number rules, regulatory costs, and daily volatility. The model thus provides regulators with a potent tool to better manage these factors in practical scenarios.

% The text of your document goes here

% This specifies the bibliography file (bib file) and the style
\bibliographystyle{plain}
\bibliography{bib}

\begin{appendices}

\section{Agent Vision} 

How actions occur are governed by code implementation details that align to assumptions both about the agents themselves and the market landscape. One crucial feature is the concept of "vision" - detailed above.  This has the effect of detailing how far an agent can access client sources of resources. In the financial industry this is considered a key feature and referred to as the "franchise value" or depth of client relationship/client base.  

Vision \(v\) is the breadth of analysis each agent can perform and ranges up to 5 unit moves in any one direction.  This is fixed for the life of an agent within a simulation.

\section{Metabolism}
\hfill \break
Let \(Met_{and}(_t, _a, _s)\) be the initial metabolism for each.
Metabolism is a metric of how much of a resource and agent uses up each turn in a simulation even if they perform no actions. This is set to be a random number at inception of a simulation. It can be thought of as the cost of capital financing that a market maker must pay internally, and or the internal resources needed to sustain a market making seat.

\section{Marginal rate of substitution: MRS}
Within \cite{Axtell} p102 - the rate at which an agent has a preference for one commodity over another is determined by the amount of each resource and the relative usage of that resource (ie "metabolism").  The ratio of such preferences is the MRS. Note that accumulation of each resource is incremented or decremented after each turn, but metabolism is a constant throughout each simulation.  Specifically this is;

\[MRS_{a,t} = \cfrac  {\cfrac{Accum_{t,a,spice}}{Met_{a,spice}}}{ \cfrac{Accum_{t,a,sugar}}{Met_{a,sugar}}   }  \]

Note that \( Met_{a,s} = c \) and does not vary through time.

\section{Welfare}
An agent's welfare function changes over time, aligning with live trading environments for government bonds. This mechanism allows for bond position accumulation, similar to actual market dynamics, creating large players ("whales") in the marketplace. Furthermore, the implementation permits participants to trade at different "prices" with other agents, eliminating the notion of a universal clearing price and promoting "welfare-improving" bilateral trade. This idea, \textbf{welfare} draws strongly from game theory and the ideas of what conditions must be met for game players to interact. In this case, welfare is defined as some relative reference of one resource to another. 

% The preference is calculated in the most basic way possible - how long until an agent (A) runs out of resource (B). Agents at each turn, have a difference welfare function broken into each commodity. 
% Cobb-Douglas functional form is used to represent the relationship between accumulation and metabolism such that: 

\[Welfare_{a,t,s_[1,2]} = {Accum_{t,a,1}} ^{\cfrac{Met_{a,1}}{Met_{a,1} + Met_{a,2}}}  \]

\section{Various simulation code bases in ABM}
Whilst not an exhaustive list, below are a series of the more common ABM simulation modeling packages available. Many of the differences between models stem from style and fluency in various programming languages. 

\begin{itemize}
\item \textbf{NetLogo}: A multi-agent programmable modeling environment.
\item \textbf{Repast}: Advanced, free, and open-source agent-based modeling and simulation platforms.
\item \textbf{MASON}: A fast discrete-event multiagent simulation library core in Java.
\item \textbf{AnyLogic}: A powerful, multi-method simulation modeling tool.
\item \textbf{GAMA}: A modeling and simulation development environment for spatially explicit agent-based simulations.
\item \textbf{AgentSheets}: A visual authoring tool for agent-based simulations.
\item \textbf{Agent-based Modelling Toolkit (ABMTk)}: A Java-based project for developing agent-based simulation models.

\item \textbf{Swarm}: A software package for multi-agent simulation of complex systems.
\item \textbf{Eclipse Modeling Framework (EMF)}: A modeling framework and code generation facility.
\item \textbf{JADE (Java Agent DEvelopment Framework)}: A software framework for developing agent applications.
\item \textbf{ABIDES (Agent-Based Interactive Discrete Event Simulation)}: A high-performance, agent-based market simulation.
\item \textbf{MAXE (Multi-Agent eXchange Ecosystem)}: An environment for developing agent-based market simulations.
\item \textbf{AdaptiveModeler}: Software for creating agent-based market simulation models for price forecasting.
\item \textbf{Gabriele}: A Java-based agent-based modeling and simulation framework.
\item \textbf{SeSAM (Semiotic-based Simulation Modeling)}: A modeling and simulation tool based on Semiotics and aimed at the design and simulation of multi-agent systems.

\end{itemize}

\end{appendices}

\end{document}